\documentclass[a4paper,fleqn,numbers]{cas-dc}
\usepackage{physics}
\usepackage{mathtools}
\usepackage{subcaption}
\usepackage{graphicx}
\usepackage{amsmath}
\usepackage{etoolbox}

\usepackage{tikz}
\usetikzlibrary{shapes,decorations.pathmorphing}
\usetikzlibrary{arrows.meta}
\usetikzlibrary{calc,patterns,angles,quotes}
\newcommand{\tikzsymbol}[2][circle]{\tikz[baseline=-0.5ex]\node[inner
	sep=2pt,shape=#1,draw,#2]{};}
\DeclareRobustCommand\dashdotted{\tikz[baseline=-0.6ex]\draw[thick,red,dashdotted] (0,0)--(0.54,0);}
\DeclareRobustCommand\dashed{\tikz[baseline=-0.6ex]\draw[thick,blue,dashed] (0,0)--(0.54,0);}
\DeclareRobustCommand\solid{\tikz[baseline=-0.6ex]\draw[thick,black] (0,0)--(0.54,0);}
\usepackage{setspace}

\usepackage[square, sort&compress, numbers]{natbib}
\usepackage{hyperref}
\hypersetup
{
  colorlinks=true,
  linkcolor=blue,
  citecolor=blue,
  urlcolor=blue,
}

\begin{document}  
\sloppy
\title[mode = title]{Kinematic analysis of multiple Compton scattering in quantum-entangled two-photon systems}   
\def\floatpagepagefraction{1}
\def\textpagefraction{.001}
\shorttitle{Compton scattering in entangled two-photon systems}
\shortauthors{Peter Caradonna}
        
\tnotetext[0]{The author gratefully acknowledge the support of the Science and Technology Facilities Council (STFC) Grant No. ST/W006383/1.}

\author[]{Peter Caradonna}
    [
        orcid=0000-0001-7689-8425
    ]

\ead{peter.caradonna@york.ac.uk}
\ead[url]{https://www.researchgate.net/profile/Peter-Caradonna}
\credit{}

\affiliation[]
    {
        organization={School of Physics, Engineering and Technology, The University of York},
       addressline={Heslington}, 
        city={YO10},
        citysep={}, 
        postcode={5DD}, 
        state={England},
        country={United Kingdom}
    }

\begin{abstract}
    \noindent
The Stokes-Mueller method is used to analyze the scattering of entangled photon pairs in a two-photon system. This study examines the scenario where one of the photons, part of a pair of maximally entangled annihilation photons, undergoes intermediate Compton scattering before both photons are detected using Compton polarimeters. The method also accounts for potential quantum-decoherence effects resulting from Compton scattering. The analysis investigates the scattering behavior in both parallel and perpendicular planes, identifying variations in the modulation factor that affect azimuthal correlations. These variations include increases, decreases, sign changes, or disappearances at certain intermediate scattering angles. This work aims to provide theoretical results that support the testing and verification of predictions made by quantum field theory.
\end{abstract}

\begin{keywords}
gamma rays\sep entanglement \sep Compton scattering
\end{keywords}
\maketitle
\section{Introduction}
Systematic investigations into the physical properties of maximally entangled annihilation photons, particularly focusing on changes after one or both photons undergo Compton scattering, are advancing rapidly. This active area of study aims to develop a deeper understanding of the behaviors and characteristics of these photons under specific experimental conditions. This interest is driven by the pivotal role these photons play in various fields, including fundamental physics~\cite{Irby2003,Gauthier2010,Moskal2024}, material science~\cite{Gauthier2010,Rozing1991}, and medical imaging~\cite{Uenomachi2022, Romanchek2023, Bass2023, Sandip2011, Kazemi2023,Hiesmayr2019}. Yet, the theoretical framework essential to support this research remains underdeveloped. 

The joint differential Compton scattering cross section for entangled annihilation photons, termed ``2-Compton'' for brevity, has been effectively resolved within the Quantum Electrodynamics (QED) framework. This analysis is applied specifically to scenarios where each photon undergoes a scattering event~\cite{Pryce1947,Snyder1948}. Bohm and Aharonov demonstrated that the probability of scattering in 2-Compton experiments of a hypothetical pair of annihilation photons, which exhibit polarization correlations similar to those of an entangled pair but are not actually entangled, would be measurably different~\cite{Bohm1957}. The primary aim of their research was to establish a lower experimental bound, allowing the initial provisional use of experimental data to demonstrate the witnessing of entanglement for the first time.

Central to ongoing research is the discovery by Bohm and Aharonov that the probability of Compton scattering in a specific direction is influenced not only by polarization but also by the entanglement between photons. This insight is crucial for developing theoretical models aimed at predicting scattering probabilities in $n$-Compton ($n>2$) scattering scenarios. Their findings underscore the necessity of incorporating the degree of entanglement into theoretical formulations. Such models are essential for accurately predicting scattering outcomes when photons undergo multiple Compton interactions. This emphasizes the importance of entanglement considerations in the foundational theory of photon interactions at quantum levels.

In the absence of a $n$-Compton scattering framework, it remains poorly understood as to whether Compton scattering effectively acts as an entanglement ``kill-switch''~\cite{Caradonna2024} or not at all~\cite{Tkachev2024}. Furthermore, there is contention that Compton scattering can serve as an alternative to a polarizer in testing the invariance of P, T, and CP in 3-gamma annihilation experiments~\cite{Moskal2024,George2024}. Intriguingly, recent experimental findings suggest that the entanglement correlations of annihilation photons could vary depending on the annihilation mechanism and molecular arrangement~\cite{Moskal2024a}.

The 3-Compton cross section is of significant importance because it involves one photon undergoing intermediate Compton scattering before both photons are detected using Compton polarimeters. Compton polarimeters belong to the category of position-sensitive devices capable of determining the orientation of the scattering plane of a Compton-scattered photon~\cite{Koide2018}. Consequently, when operating in coincidence mode, a pair of Compton polarimeters can detect entanglement-induced correlations by analyzing the coincidence count rate of the scattered annihilation photons as a function of the relative angle between the scattering planes~\cite{Kasday1975}.

Fundamentally, the 3-Compton scenario is essential for understanding how Compton scattering affects the entanglement between annihilation photons. Practically, in the context of typical PET imaging protocols~\cite{Nevo2022}, 3-Compton events account for a considerable proportion of recorded events~\cite{Ollinger1996, Barney1991}. Moreover, these events contribute to the reduced visibility of cancer growths in PET imaging~\cite{Zatcepin2023, Levin2008, Zaidi2007}.

The most immediate application of a 3-Compton model is in recently proposed quantum-entangled PET (QE-PET). This introduces a novel paradigm in which Compton polarimeters replace the scintillation detectors found in conventional PET systems (or standard segmented detector arrays are adapted for polarimetry through new analysis methods)~~\cite{Watts2021}. This technological shift is geared towards improving image quality by investigating and leveraging the quantum correlations that arise from the entanglement in the polarization between the annihilation photons~\cite{Moskal2024,Uenomachi2021,Hiesmayr2019,McNamara2014}.

The accurate treatment of the process of an intermediate scattering event involving annihilation photons is crucial for a broad range of applications. These extend beyond PET imaging to include the exploration and development of innovative instruments in scenarios involving entangled photons~\cite{Romanchek2023,Uenomachi2022,Kazemi2023}. It is important to note that typical photon detection systems at the MeV scale, such as scintillation crystals, inherently experience multiple Compton scatterings. Additionally, because of the absence of a comprehensive theory on entanglement loss, there is also a deficiency in correct modeling of event topology during the coincident detection of entangled photons.

The experimental study of the 3-Compton scenario reveals that polarization correlations remain robust at intermediate scattering angles between zero and approximately 50 degrees~\cite{Parashari2024, Strizhak2022, Ivashkin2023}. We will provide details on the calculations of the 3-Compton cross section, which have demonstrated good agreement with recent experimental findings~\cite{Bordes2024, Tkachev2024}. In the subsequent sections, we will discuss these calculations, which utilize QED embedded in a Stokes matrix formalism to model both the photon and the interacting Compton electron~\cite{Wightman1948, Fano1949, McMaster1961}.
\section{Density matrix of Compton electrons}
\label{sec:II}
Wightman~\cite{Wightman1948} introduced a novel matrix calculus that streamlined the computation of cross sections for double Compton scattering, as well as for understanding quantum correlations between entangled annihilation photons. The Wightman approach effectively encodes the quantum field theory calculations into a format that facilitates modeling of a two-photon entangled system.

Wightman represented the Klein-Nishina cross section in the form of a $4\times 4$ matrix that is averaged over the initial spin states and summed over the final spin states of the electron. Fano~\cite{Fano1949} later extended the applicability of the Compton matrix to include scattering off spin-polarized electrons. 

McMaster~\cite{McMaster1961} illustrated the extensive utility of the Fano Compton matrix and other Mueller matrices in calculating the cross sections of various polarized radiation and particles involved in different polarization-sensitive interactions. This method is shown to align with both quantum field theory calculations and experimental findings~\cite{Yoshitaka2023,Feng2020,Stock2015,Danneberg2005}.

Schmidt and Simons identified a mathematical link between the Fano density matrix method and the McMaster method, further streamlining the calculation of cross sections for annihilation photons scattering off spin-polarized electrons~\cite{Schmidt1973}. Their calculations demonstrated agreement with the experimental data, which shows that their methodology produces results consistent with experimental observations. 

Yet, the matrix method devised by the mentioned authors has not been expanded to compute cross sections for the multiple Compton scattering of entangled two-photon systems. The goal of this section is to outline a method for obtaining a density matrix that characterizes the electrons involved in the multiple scattering of a beam of photons. This scattering is caused by a finite number of independent Compton electrons that can be in a spin-polarized state configuration, as described in~\cite{Fano1949,McMaster1961}. Although the method is based on the specific scenario in which all of the struck electrons are initially at rest, the theory can encompass a broader range of situations, including those in which some or all of the electrons have non-zero momenta. This extension is achieved through the application of Lorentz transforms.

\begin{figure}[t]
 \includegraphics[width=0.49\textwidth]{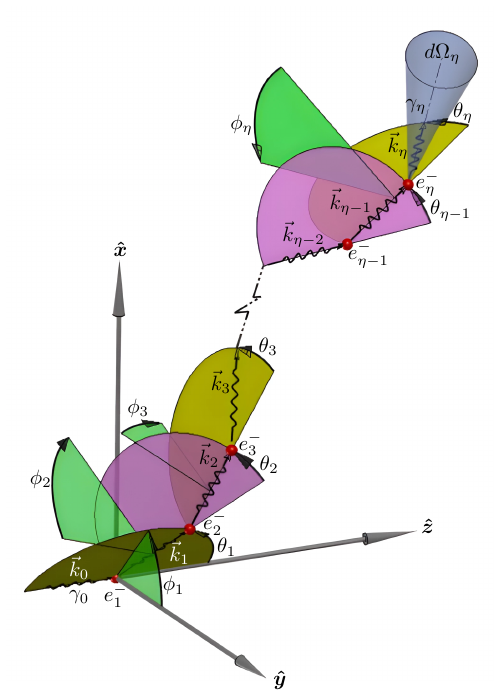}%
\caption{An arbitrary polarized photon labeled $\gamma_{0}$ represented by the Stokes vector $\ket{S}$ (see Eq.(\ref{eqn:5})) undergoing $\eta$ consecutive Compton scattering events by stationary independent electrons. The azimuthal angle $\phi_{1}$ is the angle between the scattering plane that contains the momentum vectors $\vb*{\vec{k}}_{0}$ and $\vb*{\vec{k}}_{1}$ and the $y-z$ plane. The azimuthal angle $\phi_{j}$, where $1< j\le \eta$, is the angle between the scattering plane containing $\vb*{\vec{k}}_{j-2}$ and $\vb*{\vec{k}}_{j-1}$ and the scattering plane containing $\vb*{\vec{k}}_{j-1}$ and $\vb*{\vec{k}}_{j}$. Polar angle $\theta_{1}$ is the angle between $z$-axis and $\vb*{\vec{k}}_{1}$. For $1< j\le \eta$, $\theta_{j}$ is the angle between $\vb*{\vec{k}}_{j - 1}$ and $\vb*{\vec{k}}_{j}$.}
\label{fig:1}
\end{figure}

Let $\rho^{(1,2,\cdots,\eta)}$ denote the density matrix that characterizes the distribution of $\eta$ independent Compton electrons. In the Stokes-Mueller formalism, $\rho^{(1,2,\cdots,\eta)}$ is defined in terms of Stokes parameters which are label here as $S^{(1,2,\cdots,\eta)}_{a}$ ($a=0,1,2,3$), such that
\begin{equation}
    \begin{aligned}
	&\rho^{(1,2,\cdots,\eta)} =\frac{1}{2}\times\\
	&
	\begin{bmatrix}\vspace{5.5pt}
   	S^{(1,2,\cdots,\eta)}_{0} + S^{(1,2,\cdots,\eta)}_{3} & S^{(1,2,\cdots,\eta)}_{1} - i S^{(1,2,\cdots,\eta)}_{2} \\\vspace{5.5pt}
        S^{(1,2,\cdots,\eta)}_{1} + i S^{(1,2,\cdots,\eta)}_{2} & S^{(1,2,\cdots,\eta)}_{0} - S^{(1,2,\cdots,\eta)}_{3}
	\end{bmatrix}.
    \end{aligned}
\label{eqn:1}
\end{equation}   
In Section~\ref{subsec:IIA}, a pair of $\rho^{(1,2,\cdots,\eta)}$ will be used to describe the electrons involved in the Compton scattering of a two-photon system. 

Consider a scenario in which a photon beam is initially in an arbitrary polarized state $\ket{\varphi}$ and each photon is labeled $\gamma_{0}$. This state is expanded using the basis vectors $\ket{0}$ and $\ket{1}$, such that
\begin{equation}
    \ket{\varphi}=c_{1}\ket{0}+c_{2}\ket{1},
\label{eqn:2}
\end{equation}
where the probability amplitudes $c_{1}$ and $c_{2}$ belong to the complex number field and satisfy the normalisation condition $\abs{c_{1}}^{2}$ $+$ $\abs{c_{2}}^{2}$ $=$ $1$. The basis vectors $\ket{0}$ and $\ket{1}$ correspond to the right $\ket{R}$ and left $\ket{L}$ circularly polarized basis vectors, respectively, such that
\begin{equation*}
    \ket{0} \coloneqq \ket{R} = 
    \begin{bmatrix}
		1\\
		0
    \end{bmatrix},\quad
		\ket{1} \coloneqq \ket{L} = 
    \begin{bmatrix}
		0\\
		1
    \end{bmatrix}.
\end{equation*}
The basis vectors $\ket{0}$ and $\ket{1}$ are eigenstates of the $\sigma_{3}$ Pauli matrix with eigenvalues of $\pm 1$ unit of spin angular momentum, respectively, where
\begin{equation}
    \sigma_{3}=
    \begin{bmatrix}
		1 & 0 \\
		0 & -1
    \end{bmatrix},
\label{eqn:3}
\end{equation}
and where the projection of the spin is along the trajectory of state $\ket{\varphi}$. 

The density matrix $\ket{\varphi}\bra{\varphi}$ of state $\ket{\varphi}$ is given by 
\begin{equation}
    \ket{\varphi}\bra{\varphi} =
    \begin{bmatrix}
   	\abs{c_{1}}^{2} & c_{1}^{*}c_{2} \\
	  c_{1}c_{2}^{*} & \abs{c_{2}}^{2}
    \end{bmatrix}.    
\label{eqn:4}
\end{equation}

To derive the solutions for the Stokes parameters $S^{(1,2,\cdots,\eta)}_{a}$, of the interacting electrons, it is necessary to re-express the density matrix $\ket{\varphi}\bra{\varphi}$, of Eq.(\ref{eqn:4}), in terms of its associated Stokes parameters that we label $S_{a}$. 

Stokes parameters $S_{a}$ are defined as the expectation values of the $2\times2$ unit matrix denoted by $\sigma_{0}$ and the Pauli spin matrices $\{\sigma_{1},\sigma_{2},\sigma_{3}\}$. These Stokes parameters are conveniently grouped together in the form of a four-vector represented using the bra-ket notation as $\ket{S}$, such that
\begin{equation}
    \ket{S}=
    \begin{bmatrix}\vspace{3.5pt}
		S_{0}	\\\vspace{3.5pt}
		S_{1}	\\\vspace{3.5pt}
		S_{2}	\\\vspace{3.5pt}
		S_{3}
    \end{bmatrix}= 
    \begin{bmatrix}\vspace{3.5pt}
		\bra{\varphi}\sigma_{0}\ket{\varphi}\\\vspace{3.5pt}
		\bra{\varphi}\sigma_{1}\ket{\varphi}\\\vspace{3.5pt}
		\bra{\varphi}\sigma_{2}\ket{\varphi}\\\vspace{3.5pt}
		\bra{\varphi}\sigma_{3}\ket{\varphi}
    \end{bmatrix}= 
    \begin{bmatrix}\vspace{3.5pt}
		\abs{c_{1}}^{2} + \abs{c_{2}}^{2}	\\\vspace{3.5pt}
		c_{1}c^{*}_{2} + c_{2}c^{*}_{1}	\\\vspace{3.5pt}
		i\left(c_{1}c^{*}_{2} - c_{2}c^{*}_{1}\right)\\\vspace{3.5pt}
		c_{1}c^{*}_{1}-c_{2}c^{*}_{2}
    \end{bmatrix},
\label{eqn:5}
\end{equation}
where 
\begin{equation*}
    \sigma_{0}=\begin{bmatrix}
		1 & 0 \\
		0 & 1
    \end{bmatrix},
    \quad\sigma_{1} = 
    \begin{bmatrix}
		0 & 1 \\
		1 & 0
    \end{bmatrix},
    \quad\sigma_{2}=
    \begin{bmatrix}
		0 & -i \\
		i & 0
    \end{bmatrix},
\end{equation*}
and $\sigma_{3}$ is given in Eq.(\ref{eqn:3}).

The set of four Stokes parameters represents physically measurable properties and completely characterizes the photon beam. Each component describes the fraction of intensity that passes normal through a differential solid angle element $d\Omega$ per unit of time. From the definition of the expectation values, one sees that $S_{0} = 1$ is the total normalized intensity. 
	
For a randomly polarized beam, the Stokes vector reduces to $S_{0} = 1$ and $S_{1}= S_{2}=S_{3} = 0$. The second parameter, $S_{1}$, is the difference between the degree of linear polarization in the direction parallel and perpendicular to a specified reference plane. The third, $S_{2}$, is the difference between the degree of linear polarization in directions $+45^{\circ}$ and $-45^{\circ}$, and the fourth, $S_{3}$, is the difference in the degree of circular polarization of the right and left.

By utilizing the definition of the Stokes parameters $S_{a}$, as given in Eq.(\ref{eqn:5}), it can be shown that Eq.(\ref{eqn:4}) is equivalently expressed as
\begin{equation}
    \ket{\varphi}\bra{\varphi} =\frac{1}{2}
		\begin{bmatrix}
			S_{0} + S_{3} & S_{1} - i S_{2} \\
			S_{1} + i S_{2} & S_{0} - S_{3}
		\end{bmatrix}.    
\label{eqn:6}
\end{equation}

Let $\ket{S_{\eta}}$ represent the Stokes vector associated with the final photon, denoted as $\gamma_{\eta}$, following a total of $\eta$ Compton scattering events, as shown in Fig.~\ref{fig:1}. McMaster~\cite{McMaster1961} demonstrated that in the case of multiple Compton scattering, the matrices that transform $\ket{S}\mapsto\ket{S_{\eta}}$ are of the following type:
\begin{equation}
    \begin{aligned}
    \ket{S_{\eta}}&=~T_{\eta}M_{\eta}T_{\eta-1}M_{\eta-1}\cdots T_{2}M_{2}T_{1}M_{1}\ket{S}\\
               &=~\prod_{j = \eta}^{1}T_{j}M_{j}\ket{S}.\\
    \end{aligned}
\label{eqn:7}
\end{equation}
Here $j$ is a descending integer of the product operator $\Pi$, such that $j$ runs from $j = \eta, \eta - 1, \eta - 2,\cdots, 1$. The definitions of the $4\times4$ Compton transition matrix $T_{j}$ and the rotation matrix $M_{j}$ can be found in Appendix~\ref{app:A}.

The differential cross section for detecting $\gamma_{\eta}$, as illustrated in Fig.~\ref{fig:1}, using a polarization filter characterized by a Stokes vector $\ket{D_{p}}$ (refer to Appendix \ref{app:B} for more details and examples), is given by
\begin{equation}
    \frac{d\sigma}{d\Omega_{\eta}}=~\frac{1}{2}\bra{D_{p}}\ket{S_{\eta}}=\frac{1}{2}\bra{D_{p}}\prod_{j = \eta}^{1}T_{j}M_{j}\ket{S},
\label{eqn:8}
\end{equation}
where the differential unit of solid angle $d\Omega_{\eta}$ is defined by the area it covers on the surface of a sphere centered at the point where the electron labeled $e_{\eta}^{-}$ is located. The cross section given in Eq.(\ref{eqn:8}) has already incorporated the necessary summing over the final electron spins.

However, at MeV energies, direct measurement of polarization faces experimental challenges due to the unavailability of polarization filters, unlike in the optical regime, where there is an abundant array of filters to choose from. Using a polarization-insensitive detector, we effectively measure each of the two orthogonal states of polarization and sum the results. 

Let $\ket{I}$ denote the Stokes vector for a polarization-insensitive detector, where $\ket{I} = [1,0,0,0]^{T}$ (refer to Appendix \ref{app:B} for more details). In this case, the formula for the differential cross section becomes
\begin{equation}
        \frac{d\sigma}{d\Omega_{\eta}}=\bra{I}\ket{S_{\eta}}=\bra{I}\prod_{j = \eta}^{1}T_{j}M_{j}\ket{S}.
\label{eqn:9}
\end{equation}
If the Stokes vector $\ket{S}$ and the total number of Compton scattering events $\eta$ are known, then the right-hand side of Eq.(\ref{eqn:8}) or Eq.(\ref{eqn:9}) can, in principle, be computed to determine the differential cross section.

However, the goal is to evaluate the right-hand side of Eq.(\ref{eqn:8}) or Eq.(\ref{eqn:9}) by expanding in terms of the Stokes parameters $S_{a}$. The coefficients of $S_{a}$ will be used to construct the density matrix $\rho^{(1,2,\cdots,\eta)}$, given in Eq.(\ref{eqn:1}).

To accomplish this, we use the trace method approach of Schmidt and Simons~\cite{Schmidt1973}, to calculate the differential cross section presented in Eq.(\ref{eqn:8}) or Eq.(\ref{eqn:9}) in a distinct but equivalent manner. Using $\ket{\varphi}\bra{\varphi}$ given in Eq.(\ref{eqn:6}) and $\rho^{(1,2,\cdots,\eta)}$ from Eq.(\ref{eqn:1}), the differential cross section can also be obtained by solving the expression: 
\begin{equation}
    \frac{d\sigma}{d\Omega_{\eta}}=\Tr\left[\ket{\varphi}\bra{\varphi}\cdot\rho^{(1,2,\cdots,\eta)}\right].
\label{eqn:10}
\end{equation}
Evaluating the right hand side yields
\begin{equation}
    \frac{d\sigma}{d\Omega_{\eta}}=\frac{1}{2}\sum_{a=0}^{3}S_{a}S^{(1,2,\cdots,\eta)}_{a}.
\label{eqn:11}
\end{equation}
Now, equating the right hand side of Eq.(\ref{eqn:11}) with the corresponding expression in Eq.(\ref{eqn:8}), for a polarization filter characterized by a Stokes vector $\ket{D_{p}}$, leads to
\begin{subequations}
    \begin{equation}
        \sum_{a=0}^{3}S_{ia}S^{(1,2,\cdots,\eta)}_{a} = \bra{D_{p}}\prod_{j = \eta}^{1}T_{j}M_{j}\ket{S}.
    \label{eqn:12a}
    \end{equation}
Similarly, when equating the right hand side of Eq.(\ref{eqn:11}) with that in Eq.(\ref{eqn:9}), for a polarization-insensitive detector characterized by a Stokes vector $\ket{I}$, leads to
    \begin{equation}
        \sum_{a=0}^{3}S_{a}S^{(1,2,\cdots,\eta)}_{a} = 2\bra{I}\prod_{j = \eta}^{1}T_{j}M_{j}\ket{S}.
    \label{eqn:12b}
    \end{equation}
\label{eqn:12}
\end{subequations}
By expanding both the left and right sides of equations (\ref{eqn:12a}) and (\ref{eqn:12b}) with respect to $S_{a}$, we equate the coefficients corresponding to each $S_{a}$ on both sides. Subsequently, these coefficients are solved in terms of $S^{(1,2,\cdots,\eta)}_{a}$ which are expressed in relation to the known matrix elements associated with the matrices $T_{j}$ and $M_{j}$. This process enables the determination of the density matrix given in Eq.(\ref{eqn:1}), which characterizes the distribution of $\eta$ independent Compton electrons. In Section \ref{sec:III}, we leverage Eq.(\ref{eqn:12b}) to compute the 3-Compton differential cross section for annihilation photons.
\subsection{Multiple Compton scattering of a general two-photon system}
\label{subsec:IIA}
Consider some hypothetical device the emits a two photon system defined by the density matrix $\rho$ such that
\begin{subequations}
    \begin{equation}
        \rho=\sum\limits^{N}_{s=1}p_{s}\ket{\psi_{s}}\bra{\psi_{s}},
        \label{eqn:13a}
    \end{equation}	
where $\{\ket{\psi_{s}}\}_{s=1}^{N}$ represents a set of $N$ ensemble 2-photon pure states. These states are not necessarily orthogonal, and each state is given by
\begin{equation}
   \begin{aligned}
        \ket{\psi_{s}} =&~
        c_{s1}
        \begin{bmatrix}
		1\\
		0\\
            0\\
            0\\
    \end{bmatrix}+
    c_{s2}
        \begin{bmatrix}
		0\\
		1\\
            0\\
            0\\
    \end{bmatrix}+
        c_{s3}
        \begin{bmatrix}
		0\\
		0\\
            1\\
            0\\
    \end{bmatrix}+
        c_{s4}
        \begin{bmatrix}
		0\\
		0\\
            0\\
            1\\
    \end{bmatrix}\\
    =&~c_{s1}\ket{0,0}+c_{s2}\ket{0,1}+c_{s3}\ket{1,0}+c_{s4}\ket{1,1}.\\
     \end{aligned}
     \label{eqn:13b}
\end{equation}
Alternatively, in terms of $\ket{R}$ and $\ket{L}$ notation, we can express $\ket{\psi_{s}}$ as    
\begin{equation}
      \ket{\psi_{s}}=c_{s1}\ket{R,R}+c_{s2}\ket{R,L}+c_{s3}\ket{L,R}+c_{s4}\ket{L,L},  
\end{equation}
where the probability amplitudes $c_{sr}$ ($r=1,2,3,4$) satisfy the normalization condition
\begin{equation}
   \sum_{r=1}^{4}\abs{c_{sr}}^{2}=1.
\end{equation}
The $N$ number of weights labeled $p_{s}$ satisfy the following conditions:
\begin{equation}
    \sum\limits^{N}_{s=1}p_{s} = 1;\quad 0 \le p_{s} \le1.
\end{equation}
\end{subequations}
Normalization of the weights $p_{s}$ ensures that $\Tr[\rho]=1$.

\begin{figure}[t]
\includegraphics[width=0.49\textwidth]{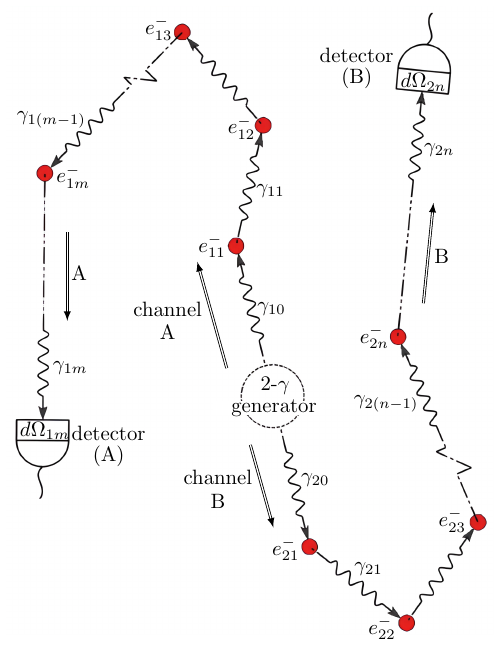}%
\caption{A schematic illustrates multiple Compton scattering events involving a pair of incident photons, designated as $\gamma_{10}$ and $\gamma_{20}$, of an arbitrary two-photon system. The photon $\gamma_{10}$ undergoes $m$ Compton scattering events through channel A, for which the $m$ electrons are defined by the density matrix $\rho^{(1,2,\cdots,m)}_{A}$. Similarly, the photon $\gamma_{20}$ undergoes $n$ Compton scattering events through channel B for which the $n$ electrons are defined by the density matrix $\rho^{(1,2,\cdots,n)}_{B}$. Subsequently, the photons exit channels A and B as $\gamma_{1m}$ and $\gamma_{2n}$, respectively, before being analyzed using two space-like separated detectors that can be characterized by the Stokes vector $\ket{D_{p}}$ or $\ket{I}$. (Refer to Appendix \ref{app:B} for more information)}
\label{fig:2}
\end{figure}

Figure (\ref{fig:2}) shows a scenario in which a pair of incident photons characterized by $\rho$ and denoted as $\gamma_{10}$ and $\gamma_{20}$ each undergo multiple Compton scattering events. Specifically, $\gamma_{10}$ Compton scatters from $m$ electrons through a region designated as channel A, while $\gamma_{20}$ Compton scatters from $n$ electrons through a region designated as channel B. Here, $m$ and $n$ are positive integers, in which $m+n$ represents the total number of scattering events by state $\rho$. The electrons resulting after the scattering events $m$ and $n$ are labeled $\gamma_{1m}$ and $\gamma_{2n}$, respectively. Let $\rho^{(1,2,\cdots,m)}_{A}$ and $\rho^{(1,2,\cdots,n)}_{B}$ denote the density matrix of the $m$ and $n$ electrons of channel A and B, respectively, where
\begin{subequations}
\begin{equation}
    \begin{aligned}
	&\rho^{(1,2,\cdots,m)}_{A} =\frac{1}{2}\times\\
	&
	\begin{bmatrix}\vspace{5.5pt}
	   	S^{(1,2,\cdots,m)}_{10} + S^{(1,2,\cdots,m)}_{13} & S^{(1,2,\cdots,m)}_{11} - i S^{(1,2,\cdots,m)}_{12} \\\vspace{5.5pt}
		  S^{(1,2,\cdots,m)}_{11} + i S^{(1,2,\cdots,m)}_{12} & S^{(1,2,\cdots,m)}_{10} - S^{(1,2,\cdots,m)}_{13}
        \end{bmatrix},
    \end{aligned}
\label{eqn:14a}
\end{equation} 
and
\begin{equation}
    \begin{aligned}
	&\rho^{(1,2,\cdots,n)}_{B} =\frac{1}{2}\times\\
	&
	\begin{bmatrix}\vspace{5.5pt}
      	S^{(1,2,\cdots,n)}_{20} + S^{(1,2,\cdots,n)}_{23} & S^{(1,2,\cdots,n)}_{21} - i S^{(1,2,\cdots,n)}_{22} \\\vspace{5.5pt}
    	  S^{(1,2,\cdots,n)}_{21} + i S^{(1,2,\cdots,n)}_{22} & S^{(1,2,\cdots,n)}_{20} - S^{(1,2\cdots,n)}_{23}
	\end{bmatrix}.
    \end{aligned}
\label{eqn:14b}
\end{equation}
\end{subequations}
If we now consider channels A and B to be space-like separated, the density matrix for the interacting electrons is given by the direct product
\begin{equation}
    \rho^{(1,2,\cdots,m)}_{A}\otimes\rho^{(1,2,\cdots,n)}_{B}.
\end{equation}

Using the trace operator, the joint differential scattering cross section of the two-photon system described by $\rho$ is then given by
\begin{equation}
    \begin{aligned}
	\frac{d^{2}\sigma}{d\Omega_{1m}d\Omega_{2n}} = \frac{1}{4}\Tr\left[\rho\cdot\rho^{(1,2,\cdots,m)}_{A}\otimes\rho^{(1,2,\cdots,n)}_{B}\right],
    \end{aligned}
\label{eqn:16}
\end{equation}
where the differential unit of solid angle $d\Omega_{1m}$ and $d\Omega_{2n}$ are defined by the elemental area they cover on the surface of a sphere centered at the point where the electrons $e_{1m}^{-}$ and $e_{2n}^{-}$, respectively, are located. By employing Eq.(\ref{eqn:12a}) or Eq.(\ref{eqn:12b}), one can derive the Stokes parameters $S^{(1,2,\cdots,m)}_{1a}$ and $S^{(1,2,\cdots,n)}_{2a}$. Consequently, the joint differential cross section, as provided in Eq.(\ref{eqn:16}), can be computed with respect to the types of polarization analyzers intended to measure $\gamma_{1m}$ and $\gamma_{2n}$. 

\section{Calculation of the 3-Compton cross section}
\label{sec:III}
\begin{figure*}[hbtp]
    \includegraphics[width=\textwidth]{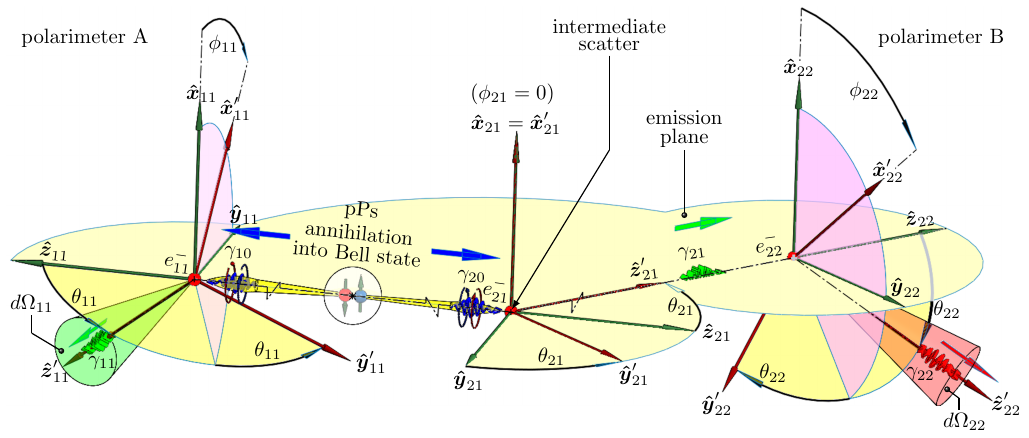}
\caption{The nomenclature for the annihilation of pPs emitting two 0.511 MeV photons, $\gamma_{10}$ and $\gamma_{20}$, in a Bell state. Initially, the photons travel in opposite directions along the unit vectors $\vb*{\hat{z}}_{11}$ and $\vb*{\hat{z}}_{21}$ ($\vb*{\hat{z}}_{11} = -\vb*{\hat{z}}_{21}$). Three local coordinate systems are defined; $\{\vb*{\hat{x}}_{ij}, \vb*{\hat{y}}_{ij}, \vb*{\hat{z}}_{ij}\}$, where $i=1,2$ (labels photon 1 or 2) and $j=1,2$ (labels the coordinate system with an electron $e_{ij}$ located at the origin). The coordinate systems are relative to an emission plane that contains the unit vectors $\{\vb*{\hat{y}}_{11},\vb*{\hat{z}}_{11}\}$, $\{\vb*{\hat{y}}_{21},\vb*{\hat{z}}_{21}\}$ and $\{\vb*{\hat{y}}_{22},\vb*{\hat{z}}_{22}\}$. Compton events occur at the origin of each coordinate system. The intermediate scattering angle $\theta_{21}$ in the coordinate system $ij=21$ is defined by $\vb*{\hat{z}}_{21}\cdot\vb*{\hat{z}}_{22}=\cos\theta_{21}$. The photon labeled $\gamma_{21}$ is the photon that results from intermediate scattering. The primed coordinate systems denoted as $\{\vb*{\hat{x}}'_{ij}, \vb*{\hat{y}}'_{ij}, \vb*{\hat{z}}'_{ij}\}$ define the coordinates of the scattered photons. The relative angle between the scattering planes of $\gamma_{11}$ and $\gamma_{22}$ is determined by the dot product $\vb*{\hat{x}}'_{11}\cdot\vb*{\hat{x}}'_{22}$ (see Eq.~(\ref{eqn:38})). We define two hypothetical Compton polarimeters, A and B, with respect to the primed coordinate systems $\{\vb*{\hat{x}}'_{11}, \vb*{\hat{y}}'_{11}, \vb*{\hat{z}}'_{11}\}$ and $\{\vb*{\hat{x}}'_{22}, \vb*{\hat{y}}'_{22}, \vb*{\hat{z}}'_{22}\}$, respectively.}
\label{fig:3}
\end{figure*}

This section outlines the derivation of the 3-Compton cross section for annihilation photons. As in the previous theoretical work for 1- and 2-Compton scattering~\cite{Snyder1948, Pryce1947, Bohm1957, Hiesmayr2019, McMaster1961} the calculation is done for a free stationary electron. We remark that the electron binding energy and Fermi motion effects for 1-, 2- (and now 3-) Compton are taken into account when the theory is implemented in simulation.~\footnote{The binding energy of the electron for light atomic nuclei is around $1\%$ of the kinetic energy of the annihilation photon and reaches $4\%$ for heavier nuclei~\cite{Zoroddu2019}. These effects are accounted for in simulation, i.e. the electron energy is reduced according to the average binding energy of the atom involved in the Compton scattering. Compton scattering is calculated in the frame where the electron is stationary and then boosted into the laboratory frame to account for the Fermi motion.} 

Hence, we can employ the general formalism from the previous section to calculate the cross section for intermediate Compton scattering by one of the annihilation photons. This calculation refers to the use of two hypothetical Compton polarimeters in a 3-Compton scattering scenario: one intermediate scattering event and one in each of the polarimeters labeled A and B, as illustrated in Fig.~\ref{fig:3}. 

Using Eq.(\ref{eqn:13b}), we set $s=1$ and label the annihilation photons $\gamma_{10}$ and $\gamma_{20}$. These photons are assumed to be produced from the disintegration of p-Ps in the ground state. The state $\ket{\psi_{1}}$ is in a Bell state described by
\begin{equation}
    \ket{\psi_{1}} = \frac{1}{\sqrt{2}}\ket{0,0}-\frac{1}{\sqrt{2}}\ket{1,1},
\label{eqn:17}
\end{equation}
where $c_{11} = 1/\sqrt{2}$, $c_{14} = -1/\sqrt{2}$, and $c_{12} = c_{13} = 0$. 

Using Eq.(\ref{eqn:13a}), the $4\times4$ density matrix $\rho$ describing state $\ket{\psi_{1}}$ takes the explicit form
\begin{equation}
    \rho = \ket{\psi_{1}}\bra{\psi_{1}}=
    \frac{1}{2}
    \begin{bmatrix}
        1 & 0 & 0 & -1 \\
	0 & 0 & 0 & 0 \\
	0 & 0 & 0 & 0 \\
	-1 & 0 & 0 & 1
    \end{bmatrix},
\label{eqn:18}
\end{equation}
where $p_{1}=1$ and $p_{s}=0$ for $s>1$.

Due to the inherent symmetries in the state $\ket{\psi_{1}}$~\cite{Bohm1957}, we are free to designate the intermediate Compton scattering event to $\gamma_{20}$ without loss of generality. This implies that $\gamma_{20}$ undergoes a double Compton scattering event.
	
Referring to Eq.(\ref{eqn:16}), the joint differential scattering cross section we wish to solve is given by 
\begin{equation}
    \frac{d^{2}\sigma}{d\Omega_{11}d\Omega_{22}} =	\frac{1}{4}\Tr\left[\rho\cdot\left(\rho^{(1)}_{A}\otimes\rho^{(1,2)}_{B}\right)\right].
\label{eqn:19}
\end{equation}
To simulate the PET environment, we will compute the cross section for the standard scenario where Compton scattering takes place from unpolarized electron spins. Note that when calculating cross sections for unpolarized electrons, a relatively simpler approach is available to obtain the solution for Eq.(\ref{eqn:19}). In this scenario, we only need to consider the upper left $3\times3$ submatrix of the $T_{j}$ and $M_{j}$ matrices~\cite{McMaster1961}. Furthermore, the Stokes parameters with subscript $a=3$ do not contribute terms to the final cross section calculation and can thus be neglected. 

Applying Eq.(\ref{eqn:14a}), the density matrix $\rho^{(1)}_{A}$, which characterizes the unpolarized electron labeled $e^{-}_{11}$ involved in the Compton scattering of $\gamma_{10}$ is given by
\begin{equation}
    \rho^{(1)}_{A} =\frac{1}{2}
    \begin{bmatrix}
	S^{(1)}_{10} & S^{(1)}_{11} - i S^{(1)}_{12} \\
	S^{(1)}_{11} + i S^{(1)}_{12} & S^{(1)}_{10}
    \end{bmatrix}.
\label{eqn:20}
\end{equation} 
Here, the subscript $1$ in the Stokes parameters $S^{(1)}_{1a}$ ($a=0,1,2$) signifies the interaction of $e^{-}_{11}$ with $\gamma_{10}$. The solutions for $S^{(1)}_{1a}$ are determined in the context of using a detector insensitive to polarization, characterized by the Stokes vector $\ket{I}$. To derive the solutions of $S^{(1)}_{1a}$, we employ Eq.(\ref{eqn:12b}) and solve for
\begin{equation}
    \sum_{a=0}^{2}S_{a}S^{(1)}_{1a} = 2\bra{I}T_{11}M_{11}\ket{S},
\label{eqn:21}
\end{equation}
where the subscripts `$11$' in the $T_{11}$ and $M_{11}$ matrices refer to the interaction matrices of the electron labeled $e^{-}_{11}$. By expanding in terms of $S_{a}$ and equating coefficients on both the left and right sides of Eq.(\ref{eqn:21}), it can be shown that
\begin{equation}
    \rho^{(1)}_{A} = \frac{1}{2}r^{2}_{0}\left(\frac{E_{11}}{E_{10}}\right)^{2}
    \begin{bmatrix}
	t^{(11)}_{11} & t^{(11)}_{12}e^{-2i\phi_{11}} \\
	t^{(11)}_{12}e^{2i\phi_{11}} & t^{(11)}_{11}
    \end{bmatrix},
\label{eqn:22}
\end{equation}
where $r_{0}$ is the classical electron radius. Furthermore, the superscript `$(11)$' of the matrix elements labeled $t^{(11)}_{mn}$ refers to the interaction terms associated with the electron $e^{-}_{11}$, and the subscripts `$mn$' refer to the position of the matrix element in the $T_{11}$ matrix. The functional form of the $t_{mn}^{(11)}$ matrix elements can be found in Appendix~\ref{app:C}.

The term $E_{10}$ represents the incident kinetic energy of $\gamma_{10}$, and in natural units, $E_{10} = 1$ ($0.511$ MeV) for annihilation photons. The term $E_{11}$ denotes the kinetic energy of $\gamma_{11}$, which is the result of the scattering of $\gamma_{10}$ by $e^{-}_{11}$. The value of this term is calculated using the Compton relation formula given by
\begin{equation}
    E_{11} = \frac{1}{2-\cos\theta_{11}};\quad \left(E_{10}=1\right).
\label{eqn:23}    
\end{equation} 
	
Using Eq.(\ref{eqn:14b}), the density matrix $\rho^{(1,2)}_{B}$, which characterizes the unpolarized electrons labeled $e^{-}_{21}$ and $e^{-}_{22}$, is given by
\begin{equation}
    \rho^{(1,2)}_{B} =\frac{1}{2}
    \begin{bmatrix}
	S^{(1,2)}_{20} & S^{(1,2)}_{21} - i S^{(1,2)}_{22} \\
	S^{(1,2)}_{21} + i S^{(1,2)}_{22} & S^{(1,2)}_{20}
    \end{bmatrix}.
\label{eqn:24}
\end{equation} 
Here, the subscript $2$ in the Stokes parameters $S^{(1,2)}_{2a}$ ($a=0,1,2$) signifies the interaction of $\gamma_{20}$ and $\gamma_{21}$ with electrons $e^{-}_{21}$ and $e^{-}_{22}$, respectively. The intermediate scattering of $\gamma_{20}\mapsto\gamma_{21}$ occurs at a polar angle of $\theta_{21}$ and an azimuthal angle of $\phi_{21}$ (set to 0 degrees), and in the case of an unpolarized electron, $S^{(1,2)}_{23} = 0$. Following this, the photon $\gamma_{21}$ undergoes another Compton scatter, $\gamma_{21}\mapsto\gamma_{22}$, inside polarimeter B, with the direction of scatter defined by the polar angle $\theta_{22}$ and the azimuthal angle $\phi_{22}$. (Refer to Fig. \ref{fig:3})
	
Applying Eq.(\ref{eqn:12b}) gives the following expression for this case:
\begin{equation}
    \sum_{a=0}^{2}\left(S_{a}S^{(1,2)}_{2a}\right) =2\bra{I}T_{22}M_{22}T_{21}M_{21}\ket{S},
\label{eqn:25}
\end{equation}
where the subscripts `$21$' and `$22$' in the $T_{21}M_{21}$ and $T_{22}M_{22}$ matrices refer to the interaction matrices of the electrons labeled $e^{-}_{21}$ and $e^{-}_{22}$, respectively.

Since $\phi_{21} = 0$, $M_{21}$ reduces to the $3\times 3$ identity matrix (refer to Appendix \ref{app:A}). As a result, the right-hand side of Eq.(\ref{eqn:25}) takes on the explicit form:
\begin{equation}
    \begin{aligned}
	&\bra{I}T_{22}M_{22}T_{21}\ket{S} = \frac{r^{4}_{0}}{4}\left(\frac{E_{22}}{E_{20}}\right)^{2}
	\begin{bmatrix}
		1 & 0 & 0 
	\end{bmatrix}\cdot\\
	&\begin{bmatrix}\vspace{5.5pt}
		t^{(22)}_{11} & t^{(22)}_{12} & 0 \\\vspace{5.5pt}
		t^{(22)}_{12} & 2 - t^{(22)}_{12} & 0 \\\vspace{5.5pt}
		0 & 0 & t^{(22)}_{33} 
	\end{bmatrix}
	\begin{bmatrix}\vspace{6.5pt}
		1 & 	0 	  & 0 \\\vspace{6.5pt}
		0 & \cos2\phi_{22} & \sin2\phi_{22} \\\vspace{6.5pt}
		0 & -\sin2\phi_{22} & \cos2\phi_{22} 
	\end{bmatrix}\\
	&\begin{bmatrix}\vspace{5.5pt}
		t^{(21)}_{11} & t^{(21)}_{12} & 0 \\\vspace{5.5pt}
		t^{(21)}_{12} & 2 - t^{(21)}_{12} & 0 \\\vspace{5.5pt}
		0 & 0 & t^{(21)}_{33} 
	\end{bmatrix}
	\begin{bmatrix}\vspace{6.5pt}
		S_{0} \\\vspace{6.5pt}
		S_{1} \\\vspace{6.5pt}
		S_{2} 
	\end{bmatrix},
    \end{aligned}
\label{eqn:26}
\end{equation}
where the superscript `$(21)$' and `$(22)$' of the $t^{(21)}_{mn}$ and $t^{(22)}_{mn}$ matrix elements refer to the interaction terms associated with the electron $e^{-}_{21}$ and $e^{-}_{22}$, respectively. The functional form of the matrix elements $t^{(2r)}_{mn}$ $(r=1,2)$ is given in Appendix \ref{app:C}. 

The term $E_{20}$ represents the incident kinetic energy of $\gamma_{20}$, and in natural units, $E_{20} =E_{10} = 1$ for the annihilation photons. The term $E_{21}$ denotes the kinetic energy of $\gamma_{21}$, which is the result of the scattering of $\gamma_{20}$ by $e^{-}_{21}$. This term is determined by the Compton relation formula given by
\begin{subequations}
    \begin{equation}
        E_{21} = \frac{1}{2-\cos\theta_{21}};\quad \left(E_{20}=1\right).
    \label{eqn:27a}    
    \end{equation}
The term $E_{22}$ denotes the kinetic energy of $\gamma_{22}$, which is the result of the scattering of $\gamma_{21}$ by $e^{-}_{22}$ in polarimeter B and is given by 
\begin{equation}
        E_{22} = \frac{E_{21}}{1+ E_{21}\left(1-\cos\theta_{22}\right)}.
    \label{eqn:27b}    
    \end{equation}
\label{eqn:27}    
\end{subequations}

After evaluating Eq.(\ref{eqn:26}) with Eq.(\ref{eqn:25}), which is expanded in terms of $S_{a}$ one can show that
\begin{equation}
    \begin{aligned}
        S^{(1,2)}_{20}S_{0} + &S^{(1,2)}_{21}S_{1} + S^{(1,2)}_{22}S_{2} =\\ 
        \frac{r_{0}^{4}}{2}\left(\frac{E_{22}}{E_{20}}\right)^{2}&\Bigg[\left(t^{(21)}_{11}t^{(22)}_{11} + t^{(21)}_{12}t^{(22)}_{12}\cos2\phi_{22}\right)S_{0}+\\
        &\left(t^{(21)}_{12}t^{(22)}_{11} + \left(2 - t^{(21)}_{12}\right)t^{(22)}_{12}\cos2\phi_{22}\right)S_{1}+\\ 
        &t^{(21)}_{33}t^{(22)}_{12}\sin 2\phi_{22}S_{2}\Bigg].
    \end{aligned}
\end{equation}
This implies that
\begin{subequations}
    \begin{equation}
        S^{(1,2)}_{20} = \frac{r^{4}_{0}}{2}\left(\frac{E_{22}}{E_{20}}\right)^{2}\left(t^{(21)}_{11}t^{(22)}_{11} + t^{(21)}_{12}t^{(22)}_{12}\cos2\phi_{22}\right),
    \label{eqn:29a}
    \end{equation}

    \begin{equation}
        \begin{aligned}
            S^{(1,2)}_{21} =~& \frac{r^{4}_{0}}{2}\left(\frac{E_{22}}{E_{20}}\right)^{2}\times\\
            &\left(t^{(21)}_{12}t^{(22)}_{11} + \left(2 - t^{(21)}_{12}\right)t^{(22)}_{12}\cos2\phi_{22}\right),
        \end{aligned}
    \label{eqn:29b}
    \end{equation}

    \begin{equation}
        \begin{aligned}
            S^{(1,2)}_{22} =~& \frac{r^{4}_{0}}{2}\left(\frac{E_{22}}{E_{20}}\right)^{2}t^{(21)}_{33}t^{(22)}_{12}\sin 2\phi_{22}\\
                           =~& \frac{r^{4}_{0}}{2}\left(\frac{E_{22}}{E_{20}}\right)^{2}2\sin\theta_{21}t^{(22)}_{12}\sin 2\phi_{22},
    \end{aligned}
    \label{eqn:29c}
    \end{equation}
\label{eqn:29}    
\end{subequations}
where we have substituted the explicit form of $t^{(21)}_{33} = 2\sin\theta_{21}$ in Eq.(\ref{eqn:29c})~\cite{Fano1949,McMaster1961}. 

By substituting the results obtained in Eq.(\ref{eqn:29}) into Eq.(\ref{eqn:24}), we determine the density matrix $\rho^{(1,2)}_{B}$. Using equations (\ref{eqn:18}), (\ref{eqn:22}) and (\ref{eqn:24}), the joint differential cross section of Eq.(\ref{eqn:19}) can be fully determined analytically and is given by
\begin{equation}
    \begin{aligned}
	&\frac{d^{2}\sigma}{d\Omega_{11}d\Omega_{22}}=\frac{r_{0}^{6}}{32}\left(\frac{E_{11}E_{22}}{E^{2}_{0}}\right)^{2}
        P\left(\theta_{11},\theta_{22},\phi_{11},\phi_{22};\theta_{21}\right),
    \end{aligned}
\label{eqn:30}
\end{equation}
where we have set $E_{10}=E_{20}=E_{0}=1$. The function $P$ can be expressed as
\begin{equation}
    \begin{aligned}
        P =&~A\left(\theta_{11},\theta_{22};\theta_{21}\right) - B\left(\theta_{11},\theta_{22};\theta_{21}\right)C\left(\phi_{11},\phi_{22};\theta_{21}\right) +\\ &D\left(\theta_{11},\theta_{22},\phi_{11},\phi_{21};\theta_{21}\right),
    \end{aligned}
\label{eqn:31}
\end{equation}
where
\begin{subequations}
    \begin{equation}
        A\left(\theta_{11},\theta_{22};\theta_{21}\right) = t^{(11)}_{11}t^{(21)}_{11}t^{(22)}_{11},
    \label{eqn:32a}
    \end{equation}
    \begin{equation}
        B\left(\theta_{11},\theta_{22}\right) = 2t^{(11)}_{12}t^{(22)}_{12},
    \label{eqn:32b}
    \end{equation}
    \begin{equation}
        \begin{aligned}
	   C\big(\phi_{11},\phi_{22};\theta_{21}\big)=&~\cos2\phi_{11}\cos2\phi_{22}-\\
                                                     &~\cos\theta_{21}\sin2\phi_{11}\sin2\phi_{22},
        \end{aligned}
    \label{eqn:32c}	
    \end{equation}
and
    \begin{equation}
        \begin{aligned}
		&D\big(\theta_{11},\theta_{22},\phi_{11},\phi_{22};\theta_{21}\big)=t^{(21)}_{12}\times\\
		&\Big[t^{(22)}_{12}\cos2\phi_{22}\left(t^{(11)}_{11}+t^{(11)}_{12}\cos2\phi_{11}\right)-t^{(11)}_{12}t^{(22)}_{11}\cos2\phi_{11}\Big],
	\end{aligned}
    \label{eqn:32d}	
    \end{equation}
\label{eqn:32}
\end{subequations}
such that the cross section is parameterized in terms of the intermediate polar angle $\theta_{21}$. The matrix elements $t^{(pq)}_{11}$ represent the average differential cross section of scattering, regardless of polarization, and therefore the function $A\left(\theta_{11},\theta_{22};\theta_{21}\right)$ represents the uncorrelated part of the 3-Compton scattering. On the other hand, the terms $t^{(pq)}_{12}$ indicate deviations from cylindrical symmetry in the scattered intensity~\cite{Fano1949}.

\section{Results}
Referring to Eq.(\ref{eqn:30}) and Appendix \ref{app:C}, in the limit $\theta_{21}\mapsto0$, we have $C\mapsto \cos2\left(\phi_{11}+\phi_{22}\right)$, $t^{(21)}_{12} \mapsto 0$, which implies that $D\mapsto0$, $t^{(21)}_{11}\mapsto 2$, and $E_{21}\mapsto E_{20}$. In this situation, Eq.~(\ref{eqn:30}) reduces to
\begin{equation}
    \begin{aligned}
        &\left(\frac{r_{0}^{2}}{2}\right)\left(\frac{r_{0}^{4}}{16}\right)\left(\frac{E_{11}}{E_{0}}\right)^{2}\left(\frac{E_{22}}{E_{0}}\right)^{2}\times\\
	&\bigg[t^{(11)}_{11}t^{(22)}_{11} - t^{(11)}_{12}t^{(22)}_{12}\cos2\left(\phi_{11}+\phi_{22}\right)\bigg],\quad\mbox{for $\theta_{21}=0$}.
    \end{aligned}
\label{eqn:33}
\end{equation}
With the exception of a factor of $r_{0}^{2}/2$, this result precisely corresponds to the well-known Pryce-Ward cross section for 2-Compton scattering of annihilation photons~\cite{Pryce1947}.

To analyze the 3-Compton cross section given in Eq.(\ref{eqn:30}) in the general case where the intermediate scattering angle $0\le\theta_{21}\le\pi$, we simplify the analysis by setting $\theta_{11}=\theta_{22}=\theta$. This choice restricts the analysis to a subset of possible scattering geometries. With this condition, we can simplify the notation as follows: $A\left(\theta_{11},\theta_{22};\theta_{21}\right)\mapsto A\left(\theta;\theta_{21}\right)$, $B\left(\theta_{11},\theta_{22}\right)\mapsto B\left(\theta\right)$, and $D\big(\theta_{11},\theta_{22},\phi_{11},\phi_{22};\theta_{21}\big)\mapsto D\big(\theta,\phi_{11},\phi_{22};\theta_{21}\big)$, while the function $C\big(\phi_{11},\phi_{22};\theta_{21}\big)$ remains unchanged.
	
Furthermore, the forthcoming analysis will involve ratios of quantities. In these cases, factors other than the function $P$ in Eq.(\ref{eqn:30}) cancel out when the ratios are taken. For this reason, only the function $P$ given in Eq.(\ref{eqn:31}) is considered.
	
We examine four scattering scenarios depicted in Fig.~\ref{fig:4}, each varying in the orientation of the scattering planes of $\gamma_{11}$ and $\gamma_{22}$ with respect to the emission plane. 

\begin{subequations}
In Fig.~\ref{fig:4a}, both the scattering planes of $\gamma_{11}$ and $\gamma_{22}$ lie within the emission plane ($\phi_{11} = \phi_{22} = 0$). Let $P^{\parallel}\left(\theta;\theta_{21}\right)$ define this probability such that
    \begin{equation}
        \begin{aligned}
            &P^{\parallel}\left(\theta;\theta_{21}\right) = \\
            &A\left(\theta;\theta_{21}\right) - B\left(\theta\right)C\left(0,0;\theta_{21}\right) + D\left(\theta,0,0;\theta_{21}\right).
        \end{aligned}
    \end{equation}
In Fig.~\ref{fig:4b}, the scattering plane of $\gamma_{11}$ is perpendicular to the emission plane ($\phi_{11} = \pi/2$), while the scattering plane of $\gamma_{22}$ is in the emission plane ($\phi_{22} = 0$). Let $P_{11}^{\perp}\left(\theta;\theta_{21}\right)$ define this probability such that
    \begin{equation}
        \begin{aligned}
            &P_{11}^{\perp}\left(\theta;\theta_{21}\right) = \\
            &A\left(\theta;\theta_{21}\right) - B\left(\theta\right)C\left(\frac{\pi}{2},0;\theta_{21}\right) + D\left(\theta,\frac{\pi}               {2},0;\theta_{21}\right).
        \end{aligned}
    \end{equation}
\begin{figure}[]
    \centering
    \begin{subfigure}{0.48\textwidth}
    \includegraphics[width=0.98\linewidth]{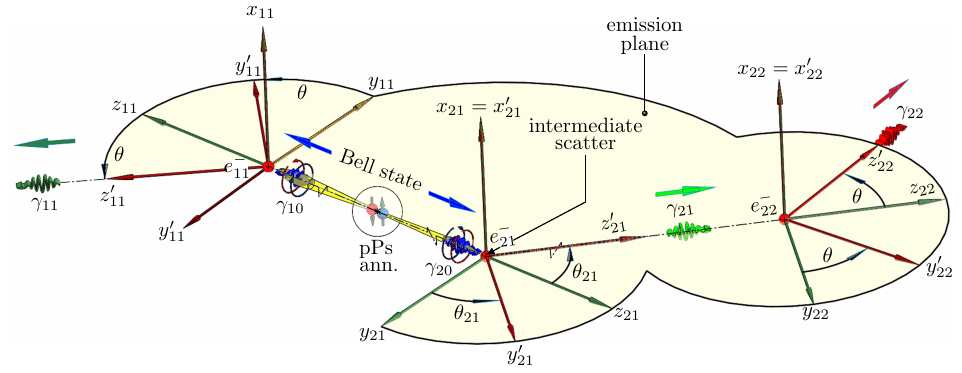}
        \caption{Scattering geometry with $\gamma_{11}$ and $\gamma_{22}$ scattering in the emission plane, represented by the probability function $P^{\parallel}\left(\theta;\theta_{21}\right)$.}
        \label{fig:4a}
    \end{subfigure}
		
    \begin{subfigure}{0.48\textwidth}
    \includegraphics[width=0.98\textwidth]{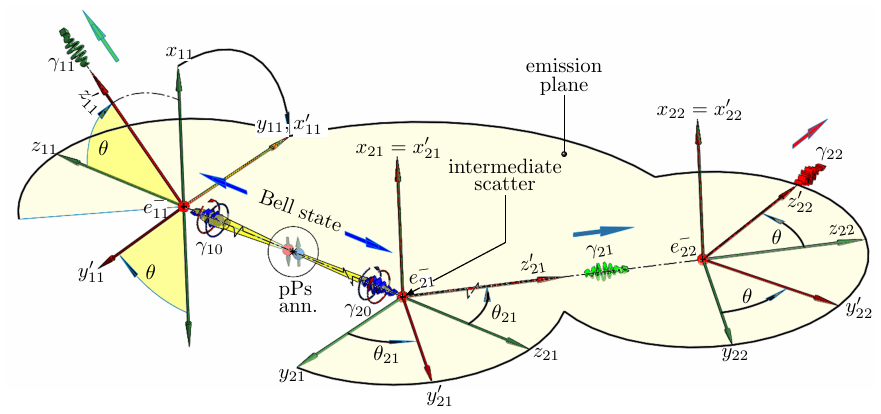}
    \caption{Scattering geometry with $\gamma_{11}$ scattering perpendicular to the emission plane and $\gamma_{22}$ scattering in the emission plane, described by the probability function $P_{11}^{\perp}\left(\theta;\theta_{21}\right)$.}
    \label{fig:4b}
    \end{subfigure}
    \begin{subfigure}{0.48\textwidth}
    \includegraphics[width=0.98\textwidth]{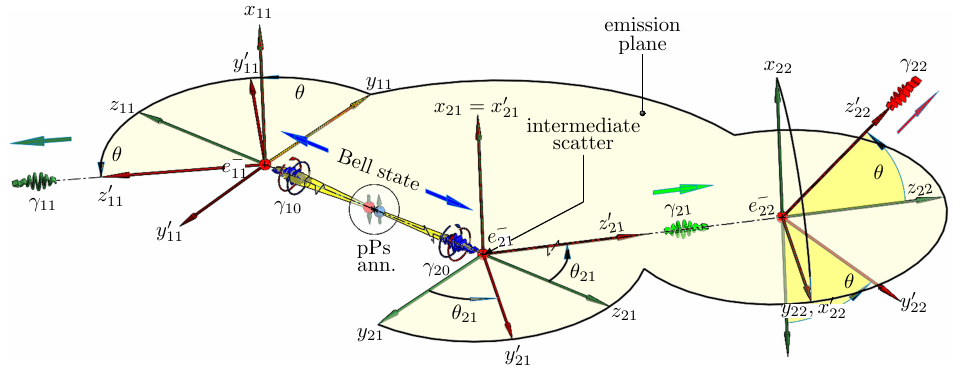}
    \caption{Scattering geometry with $\gamma_{11}$ scattering in the emission plane and $\gamma_{22}$ scattering perpendicular to the emission plane, described by the probability function $P_{22}^{\perp}\left(\theta;\theta_{21}\right)$.}
    \label{fig:4c}
    \end{subfigure}
    \begin{subfigure}{0.48\textwidth}
        \includegraphics[width=0.98\textwidth]{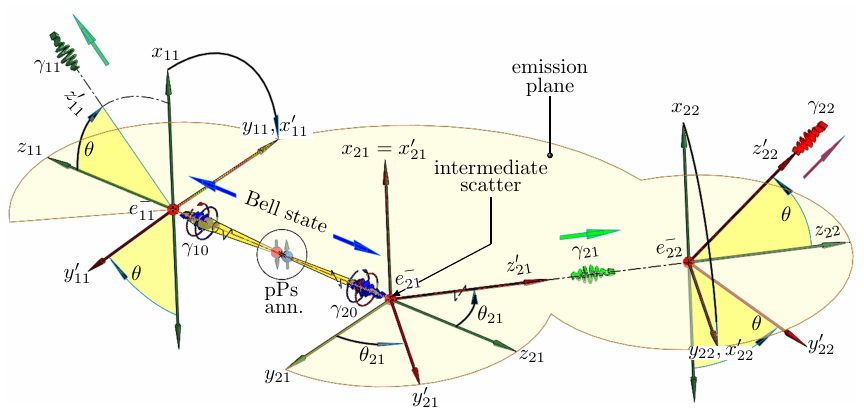}
	\caption{Scattering geometry with $\gamma_{11}$ and $\gamma_{22}$ scattering perpendicular to the emission plane, described by the                 probability function $P_{11,22}^{\perp}\left(\theta;\theta_{21}\right)$.}
	\label{fig:4d}
    \end{subfigure}
\caption{The four analyzed scattering patterns depict the scattering of $\gamma_{11}$ and $\gamma_{22}$ either within or perpendicular to the      emission plane.}
\label{fig:4}
\end{figure}
    
\begin{figure*}[hbtp]
    \centering
    \includegraphics[width=1.0\textwidth]{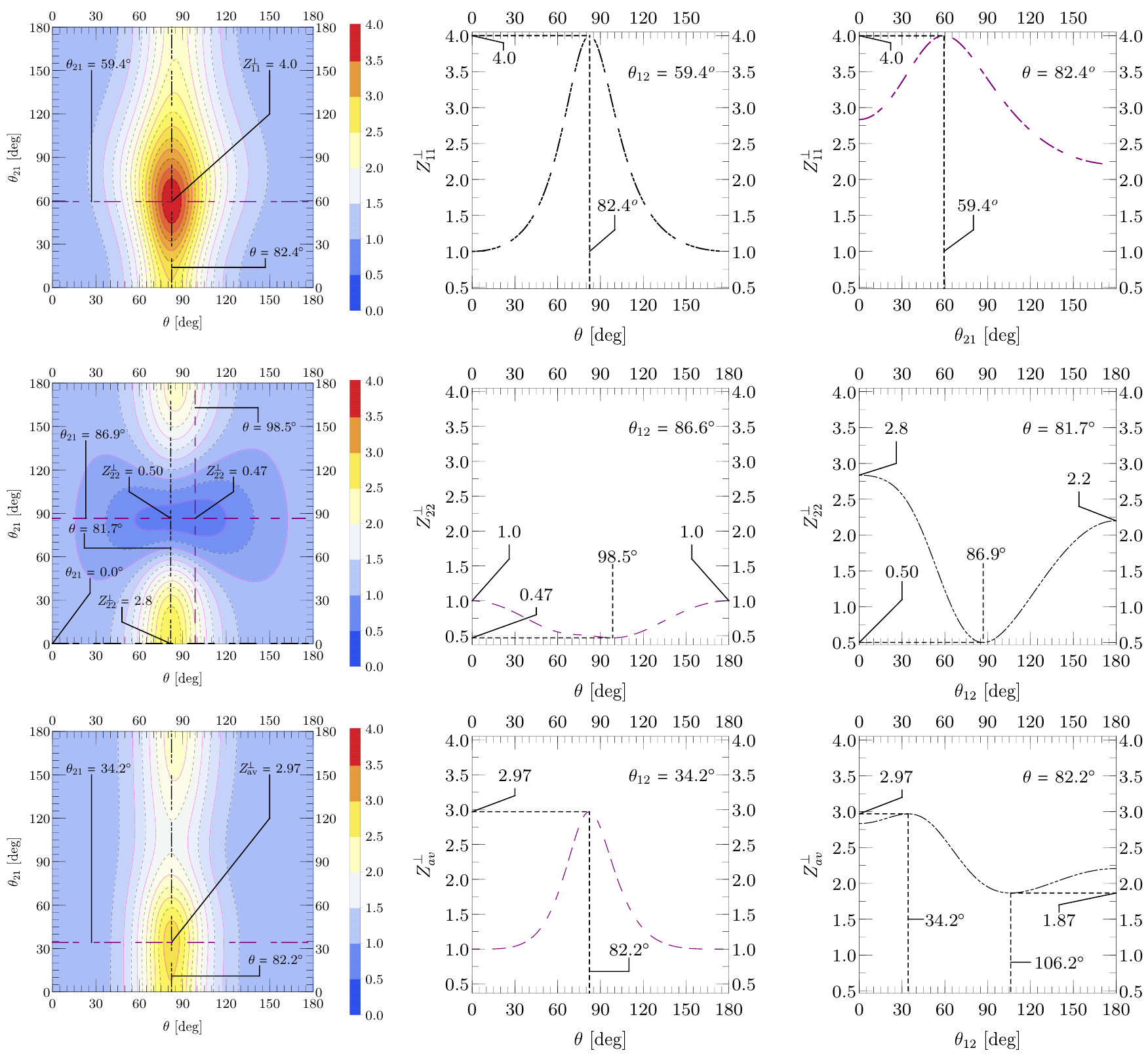} 
    \caption{The counting ratios as given in Eq.(\ref{eqn:35}). Top row: The counting ratio $Z^{\perp}_{11}\left(\theta;\theta_{21}\right)$ (c.f. Eq.(\ref{eqn:35a})) is obtained by dividing the counting rates from the scattering scenario given in Fig.~\ref{fig:4b} by those in Fig.~\ref{fig:4a}. Middle row: The counting ratio $Z^{\perp}_{22}\left(\theta;\theta_{21}\right)$  (c.f. Eq.(\ref{eqn:35b})) is obtained by dividing the counting rates from the scattering scenario shown in Fig.~\ref{fig:4c} by those in Fig.~\ref{fig:4a}. Bottom row: The average of the counting ratio $Z^{\perp}_{av}\left(\theta;\theta_{21}\right)$ (c.f. Eq.(\ref{eqn:35c})) achieves its maximum and minimum values of $2.97$ and $1.87$ at $\big(\theta = 82.2^{o}, \theta_{21} = 34.2^{o}\big)$ and $\big(\theta = 82.2^{o}, \theta_{21} = 106.2^{o}\big)$, respectively.}
\label{fig:5}
\end{figure*}
		
In Fig.~\ref{fig:4c}, the scattering plane of $\gamma_{11}$ is parallel to the emission plane ($\phi_{11} = 0$), while the scattering plane of $\gamma_{22}$ is perpendicular to the emission plane ($\phi_{22} = \pi/2$). Let $P_{22}^{\perp}\left(\theta;\theta_{21}\right)$ define this probability such that	
    \begin{equation}
        \begin{aligned}
            &P_{22}^{\perp}\left(\theta;\theta_{21}\right) = \\
		&A\left(\theta;\theta_{21}\right) - B\left(\theta\right)C\left(0,\frac{\pi}{2};\theta_{21}\right) + D\left(\theta,0,\frac{\pi}                  {2};\theta_{21}\right).
	\end{aligned}
	\end{equation}
Finally, in Fig.~\ref{fig:4d}, the scattering planes of $\gamma_{11}$ and $\gamma_{22}$ are perpendicular to the emission plane ($\phi_{11} = \phi_{22} = \pi/2$). Let $P_{(11,22)}^{\perp}\left(\theta;\theta_{21}\right)$ define this probability such that
    \begin{equation}
        \begin{aligned}
            &P_{(11,22)}^{\perp}\left(\theta;\theta_{21}\right) = \\
		&A\left(\theta;\theta_{21}\right) - B\left(\theta\right)C\left(\frac{\pi}{2},\frac{\pi}{2};\theta_{21}\right) + D\left(\theta,\frac{\pi}        {2},\frac{\pi}{2};\theta_{21}\right).
        \end{aligned}
    \end{equation}
    \label{eqn:34}
\end{subequations}
From the probabilities defined in Eq.(\ref{eqn:34}), we calculate the following expected counting ratios, denoted as $Z$:
\begin{subequations}
    \begin{equation}
        Z^{\perp}_{11} = \frac{P_{11}^{\perp}\left(\theta;\theta_{21}\right)}{P^{\parallel}\left(\theta;\theta_{21}\right)},
    \label{eqn:35a}
    \end{equation}
    \begin{equation}
        Z^{\perp}_{22} = \frac{P_{22}^{\perp}\left(\theta;\theta_{21}\right)}{P^{\parallel}\left(\theta;\theta_{21}\right)},
    \label{eqn:35b}
    \end{equation}
and we define the average of the counting ratios given by $Z^{\perp}_{av}$ such that
    \begin{equation}
        Z^{\perp}_{av} = \frac{1}{2}\left(Z^{\perp}_{11} + Z^{\perp}_{22}\right).
    \label{eqn:35c}
    \end{equation}
\label{eqn:35}
\end{subequations}
Equations (\ref{eqn:35a}) and (\ref{eqn:35b}) represent the expected ratios for an ideal experiment using the scattering geometries shown in Fig.~\ref{fig:4b} and Fig.~\ref{fig:4c}, respectively, normalized by the counting rate defined by the scattering geometry shown in Fig.~\ref{fig:4a}. These ratios are plotted in Fig.~\ref{fig:5} as a function of both $\theta$ and $\theta_{21}$ and for specific values of $\theta$ and $\theta_{21}$. 
	
\begin{figure}[t]
    \centering
    \includegraphics[width=0.49\textwidth]{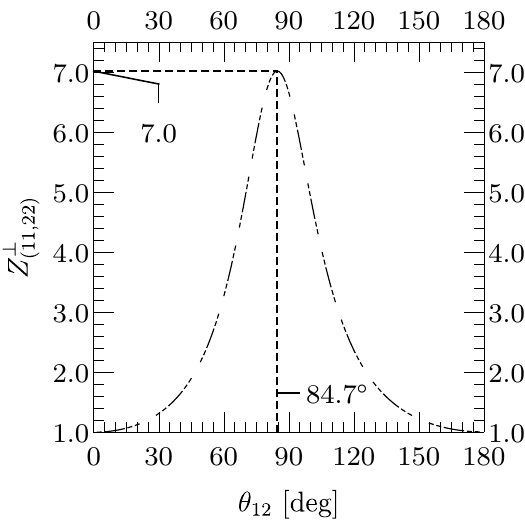}%
\caption{The counting ratio $Z_{(11,22)}^{\perp} = Z^{\perp}_{11}$ $/$ $Z^{\perp}_{22}$ is obtained by evaluating $Z^{\perp}_{11}$ at $\theta=82.4^{o}$ (refer to top row, far right plot in Fig.(\ref{fig:5})) and $Z^{\perp}_{22}$ at $\theta=81.7^{o}$ (refer to middle row, far right plot in Fig.(\ref{fig:5})).}
\label{fig:6}
\end{figure}

Furthermore, we define the ratio $Z_{(11,22)}^{\perp}$ as
\begin{equation}
    Z_{(11,22)}^{\perp} = \frac{Z^{\perp}_{11}}{Z^{\perp}_{22}} = \frac{P^{\perp}_{11}\left(\theta = 82.4^{o};\theta_{21}\right)}{P^{\perp}_{22}\left(\theta = 81.7^{o};\theta_{21}\right)},
\label{eqn:36}
\end{equation}
The ratio $Z_{(11,22)}^{\perp}$ is depicted in Fig.~\ref{fig:6} as a function of the intermediate scattering angle $\theta_{21}$. The numerator and denominator of this ratio are calculated at the corresponding scattering angles $\theta = 82.4^{o}$ (refer to top row, far right plot in Fig.(\ref{fig:5})) and $\theta = 81.7^{o}$ (refer to middle row, far right plot in Fig.(\ref{fig:5})), respectively, where the functions $Z^{\perp}_{11}$ and $Z^{\perp}_{22}$ achieve their maximum value. In particular, at an intermediate scattering angle of $\theta_{21}=84.7^{o}$, the frequency of detecting $\gamma_{11}$ scattering at right angles to the emission plane (Fig.~\ref{fig:4b}) is expected to be approximately seven times greater compared to the scenario where $\gamma_{22}$ scatters at right angles to the emission plane (Fig.~\ref{fig:4c}).
	
We have additionally examined the correlations in the coincidence counting rates of $\gamma_{11}$ and $\gamma_{22}$ as a function of the relative angle between the normal of each scattering plane defined by the unit vectors $\vb*{\hat{x}}'_{11}$ and $\vb*{\hat{x}}'_{22}$, respectively. 
	
Using Eq.(\ref{eqn:31}), we define an azimuthal correlation function for 3-Compton scattering, denoted as $R$ such that
\begin{subequations}
    \begin{equation}
        R\left(\phi_{11},\phi_{22};\theta_{11},\theta_{21},\theta_{22}\right) = \frac{P}{A}=\alpha-\beta C,
    \label{eqn:37a}
    \end{equation}
where
    \begin{equation}
        \alpha=1+\frac{D}{A}\quad\mbox{and}\quad\beta=\frac{B}{A}.
    \label{eqn:37b}
    \end{equation}
\label{eqn:37}
\end{subequations}
	
The dependent variable of $R$ is characterized by the azimuthal angles $\phi_{11}$ and $\phi_{22}$, and parameterized with respect to the Compton scattering angles $\theta_{11}$, $\theta_{21}$ and $\theta_{22}$.
	
In the general case, the scalar product of the normal to each scattering plane can be characterized as
\begin{equation}
    \vb*{\hat{x}}'_{11}\cdot\vb*{\hat{x}}'_{22} = \cos\phi_{11}\cos\phi_{22} - \cos\theta_{21}\sin\phi_{11}\sin\phi_{22}.
\label{eqn:38}
\end{equation}
We examine these correlations under two distinct scenarios: In the first scenario and referring to Eq.(\ref{eqn:37a}) and Eq.(\ref{eqn:32c}), we set $\phi_{22}=0$ and find that
\begin{figure}[t]
    \includegraphics[width=0.49\textwidth]{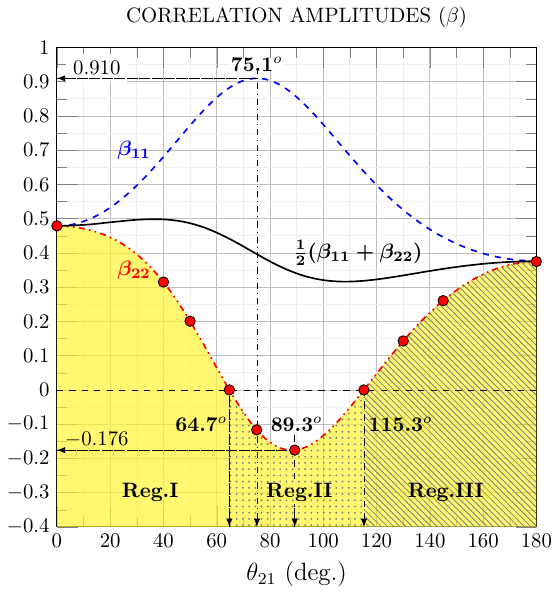}%
\caption{The correlation amplitudes evaluated for $\gamma_{11}$ and $\gamma_{22}$ scattering at specified angles of $\theta_{11}=$ $\theta_{22}=$ $\theta$. In the case of $\mathcal{\beta}_{11}$ (\dashed), we set $\phi_{22}=0$, $-\pi\le\phi_{11}\le\pi$ and evaluated for specific Compton scattering angle of $\theta = 82.4^{o}$ in both polarimeters. In the case of $\mathcal{\beta}_{22}$ (\dashdotted), we set $\phi_{11}=0$, $-\pi\le\phi_{22}\le\pi$ and evaluated for specific Compton scattering angle of $\theta = 81.7^{o}$ in both polarimeters. These results are shown in conjunction with their average $\frac{1}{2}(\mathcal{\beta}_{11} + \mathcal{\beta}_{22})$ (\solid). The points marked by red circles (\tikzsymbol{mark size=2.0pt, mark= o,fill=red}), on specific regions on the curve $\beta_{22}$, correspond to instances where the behavior of the function $R_{22}(\phi_{22})$ is visually demonstrated in Fig.~\ref{fig:8}.}
\label{fig:7}
\end{figure}
	
\begin{equation}
    R_{11}\left(\phi_{11};\theta_{11},\theta_{21},\theta_{22}\right) =\alpha_{11} -\beta_{11}\cos2\phi_{11}.
\label{eqn:39}
\end{equation}
where 
\begin{subequations}
    \begin{equation}
        \alpha_{11}\left(\theta_{21},\theta_{22}\right) =1+\frac{t^{(21)}_{12}t^{(22)}_{12}}{t^{(21)}_{11}t^{(22)}_{11}},
    \label{eqn:40a}
    \end{equation}
and the corresponding correlation amplitude is given by
    \begin{equation}
        \beta_{11}\left(\theta_{11},\theta_{21},\theta_{22}\right) =
        \frac{t^{(11)}_{12}\left[2 t^{(22)}_{12} + t^{(21)}_{12}\left(t^{(22)}_{11} - t^{(22)}_{12}\right)\right]}{t^{(11)}_{11} t^{(21)}_{11} t^{(22)}_{11}}.
    \label{eqn:40b}
    \end{equation}
\label{eqn:40}	
\end{subequations}
Similarly in the second scenario, we set $\phi_{11}=0$, and find that
\begin{equation}
    R_{22}\left(\phi_{22};\theta_{11},\theta_{21},\theta_{22}\right) =\alpha_{22} -\beta_{22}\cos2\phi_{22}.
\label{eqn:41}	
\end{equation}
where
\begin{subequations}
    \begin{equation}
        \alpha_{22}\left(\theta_{11},\theta_{21}\right) =1 - \frac{t^{(11)}_{12}t^{(21)}_{12}}{t^{(11)}_{11}t^{(21)}_{11}},
    \label{eqn:42a}	
    \end{equation}
and the corresponding correlation amplitude is given by
    \begin{equation}
        \beta_{22}\left(\theta_{11},\theta_{21},\theta_{22}\right) =
        \frac{t^{(22)}_{12}\left[t^{(11)}_{12}\left(2 - t^{(21)}_{12}\right) - t^{(11)}_{11}t^{(21)}_{12}\right]}{t^{(11)}_{11} t^{(21)}_{11} t^{(22)}_{11}}.
    \label{eqn:42b}	
    \end{equation}
\label{eqn:42}		
\end{subequations}
\begin{figure}[t]
    \centering
        \includegraphics[width=0.49\textwidth]{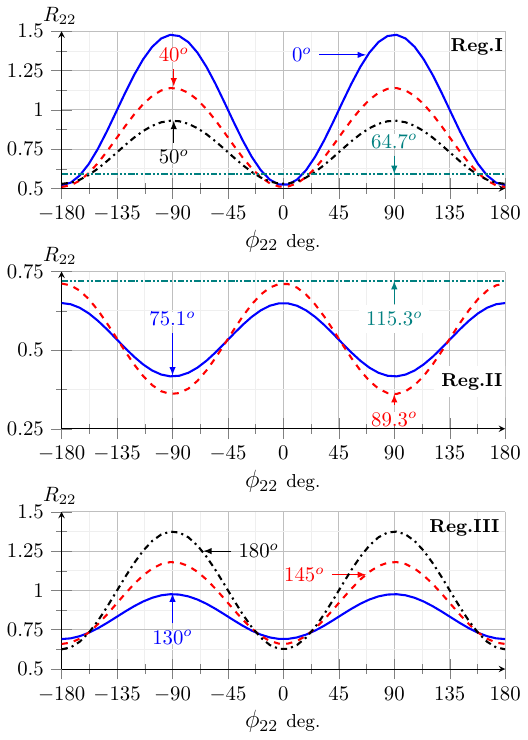}%
    \caption{The function $R_{22}\left(\phi_{22};\theta_{21}\right)$ is evaluated for $\gamma_{11}$ and $\gamma_{22}$ scattering at a fixed angle of $\theta_{11}=$ $\theta_{22}=$ $\theta=81.7^{o}$. The plots are calculated for various values of the intermediate scattering angle $\theta_{21}$ indicated by solid red circles along the plot of $\beta_{22}$ plotted in Fig. \ref{fig:7}. In particular, a sign change in the correlation amplitude is observed in Region II. Furthermore, no azimuthal correlations are predicted at two specific intermediate angles, namely $\theta_{21}=64.7^{o}$ and $115.3^{o}$. The values of $\alpha_{22}$ and its corresponding correlation amplitude $\beta_{22}$ for these graphs are provided in the Tab. \ref{tab:1}.}
\label{fig:8}
\end{figure}
In these specific scenarios, the scattering geometry resembles those illustrated in Fig.~\ref{fig:4b} and Fig.~\ref{fig:4c} respectively. However, the distinction lies in the fact that in the first scenario (when $\phi_{22}=0$), the azimuthal angle $\phi_{11}$ of $\gamma_{11}$ can assume any value between $-\pi\le\phi_{11}\le\pi$. Similarly, in the second scenario (when $\phi_{11}=0$), the azimuthal angle $\phi_{22}$ of $\gamma_{22}$ is also unrestricted and can vary between $-\pi\le\phi_{22}\le\pi$. In these two particular scattering scenarios, the angle $\phi_{11}$ or $\phi_{22}$ serves as a direct indicator of the relative angle between the scattering planes, since in the first scenario we have $\vb*{\hat{x}}'_{11}\cdot\vb*{\hat{x}}'_{22}=\cos\phi_{11}$ and in the second we have $\vb*{\hat{x}}'_{11}\cdot\vb*{\hat{x}}'_{22}=\cos\phi_{22}$ (cf. evaluating Eq.(\ref{eqn:38})).
\begin{table}[width=1.0\linewidth,cols=7,pos=h]
\caption{In the context of an ideal experiment, the numerical values are provided for $\alpha_{22}$ and their associated correlation amplitudes $\beta_{22}$. These theoretical values correspond to specific instances where the behavior of the function $R_{22}$ is visually demonstrated in Fig.~\ref{fig:8}.}\label{tab:1}
\begin{tabular*}{\tblwidth}{@{} c|cc|cc|cc@{} }
   \toprule
   \multicolumn{7}{c} {$R_{22}\left(\phi_{22};\theta_{21},\theta=81.7^{o}\right) = \alpha_{22}-\beta_{22}\cos2\phi_{22}$}\\
    \midrule
    &\multicolumn{2}{c|}{Region I}&\multicolumn{2}{c|}{Region II}&\multicolumn{2}{c}{Region III}\\
	\midrule
	$\theta_{21}$ (deg.)& $\alpha_{22}$ & $\beta_{22}$ & $\alpha_{22}$ & $\beta_{22}$ & $\alpha_{22}$ & $\beta_{22}$ \\
	\midrule
	$0$ 		& $1.000$ 	& $0.479$ 	&  			&  			&  			&			\\
	$40$		& $0.825$ 	& $0.315$ 	&  			&  			&  			&			\\
	$50$		& $0.731$ 	& $0.201$	&  			&  			&  			&			\\
	$64.734$ 	& $0.593$	&  $0$ 		& 	 		&  			& 			&			\\
	$75$		& 			&  			& $0.533$	& $-0.117$	&  			&			\\
	$89.291$ 	& 			& 			& $0.536$	& $-0.176$	& 			&			\\
	$115.266$	&  			&  			& $0.720$	& $0$		& 			& 			\\
	$130$		&  			&  			&  			&			& $0.833$	&  $0.143$ \\
	$145$		&  			&  			&  			&			& $0.920$	&  $0.261$ \\
	$180$		&  			&  			&  			&			& $1.000$	&  $0.375$ \\
\bottomrule
\end{tabular*}
\end{table}
 
The plots of correlation amplitudes $\beta_{11}$ and $\beta_{22}$, along with their average $\frac{1}{2}(\mathcal{\beta}_{11} + \mathcal{\beta}_{22})$, as functions of the intermediate scattering angle $\theta_{21}$ are shown in Fig.~\ref{fig:7}. These plots correspond to the scenario in which the scattering angles $\theta_{11}$ and $\theta_{22}$ are set to $\theta$. Specifically, $\beta_{11}$ is evaluated at $\theta = 82.4^{o}$, while $\beta_{22}$ is evaluated at $\theta = 81.7^{o}$.
	
An observation can be made in Fig.~\ref{fig:7} about the behavior of $\beta_{11}$ and $\beta_{22}$ as the intermediate scattering angle $\theta_{21}$ increases from zero degrees. Initially, these correlation amplitudes diverge. However, they eventually converge, although at a lower value compared to $\theta_{21}=0^{o}$ (see Table \ref{tab:1} for tabulated values). This convergence phenomenon is notable because it occurs when $\gamma_{11}$ and $\gamma_{22}$ experience a backscatter at $\theta_{21} = 180^{o}$. Furthermore, it should be noted that initially $\beta_{11}$ increases monotonically as the intermediate scattering angle increases. It reaches a maximum value of $0.910$ around $\theta_{21} = 75.1^{o}$ before monotonically decreasing. In particular, the sign of $\beta_{11}$ remains positive throughout.
	
The correlation amplitude $\beta_{22}$ exhibits varying properties in different regions of the intermediate scattering angle $\theta_{21}$. In Region I ($0\le\theta_{21}<64.7^{o}$), $\beta_{22}$ maintains a positive sign. However, in Region II ($64.7^{o}<\theta_{21}<115.3^{o}$), $\beta_{22}$ changes sign and becomes negative, and in Region III ($115.3^{o}<\theta_{21}\le180^{o}$), $\beta_{22}$ reverts to a positive sign, indicating a return to a scattering behavior similar to that in Region I.

Furthermore, $\beta_{22}$ reaches a minimum value of $-0.176$ at $\theta_{21} = 89.3^\circ$ in Region II. Above $89.3^{o}$ degrees, $\beta_{22}$ begins to increase. At approximately $\theta_{21} = 64.7^{o}$ and $\theta_{21} = 115.3^{o}$, the correlation amplitude becomes zero, indicating the absence of azimuthal correlations between $\gamma_{11}$ and $\gamma_{22}$. 
	
Figure~\ref{fig:8} shows the behavior of $R_{22}$ (Eq.~(\ref{eqn:41})) as a function of the relative angle between the scattering planes, defined by $\phi_{22}$, given in degrees, for selected values of $\theta_{21}$ highlighted as large solid red circles shown in Fig.~{\ref{fig:7}}. Furthermore, the associated values of $\alpha_{22}$ and $\beta_{22}$ of $R_{22}$ can be found in Table~\ref{tab:1}. 
 
\section{Conclusion}
We have developed a novel matrix framework to calculate the 3-Compton scattering cross section for entangled annihilation photons, considering potential entanglement degradation due to Compton scattering. Our analysis of the azimuthal correlations in various kinematic regions for the scattered photons shows significant variations, identifying areas where future experimental investigations of 3-Compton could be targeted to examine the core principles of the theory.

The flexibility of this framework allows for its integration into Monte Carlo particle simulations. It is adaptable, enabling the inclusion of other interaction processes and providing a basis for exploring entanglement and decoherence at energy scales, such as the GeV scale, that have not yet been fully explored.

This work establishes a foundation for a more detailed theoretical understanding of the scattering dynamics of entangled photons at the MeV scale. These processes are important for accurate modeling of annihilation photons in PET, analyzing event topology for correlated detection of entangled quanta, and improving the precision of fundamental tests at the MeV scale.

\appendix
\section{Compton transition matrices}
\label{app:A}
The Compton transition matrix $T_{pj}$ is given by
\begin{equation}
	T_{pj}=\frac{r_{0}^{2}}{2}\left(\frac{E_{pj}}{E_{p(j-1)}}\right)^{2}
    \begin{bmatrix}\vspace{5.5pt}
	t^{(pj)}_{11} & t^{(pj)}_{12}        & 0             & t^{(pj)}_{14} \\\vspace{5.5pt}
	t^{(pj)}_{12} & 2 - t^{(pj)}_{12}    & 0             & t^{(pj)}_{24} \\\vspace{5.5pt}
	0            & 0                   & t^{(pj)}_{33}  & t^{(pj)}_{34} \\\vspace{5.5pt}
	t^{(pj)}_{41} & t^{(pj)}_{42}        & t^{(pj)}_{43}  & t^{(pj)}_{44}
    \end{bmatrix},
	\label{eqn:A1}
\end{equation}
where $1 \le j \le \eta$ and $\eta$ denotes the total number of Compton scattering events and $r_{0}$ represents the classical electron radius. The subscript $p=1$ or $p=2$ labels the photon within the 2-photon system. The functional form of the $t^{(pj)}_{mn}$ matrix elements are given in Table \ref{tab:2}, with further information available in~\cite{Fano1949,McMaster1961}.

$E_{p(j-1)}$ defines the kinetic energy of $\gamma_{p(j-1)}$ after the $(j-1)$-th scattering event and $E_{pj}$ defines the kinetic energy of $\gamma_{pj}$ after the $j$-th scattering event, traveling in a direction defined by the angles $(\phi_{pj},\theta_{pj})$. We work in units where $\hbar, c = 1$ and measure the kinetic energies of the photons in units of $mc^{2} = 0.511$ MeV. 

When $j=1$, which implies $j-1=0$, then $E_{p0}$ represents the incident energy of $\gamma_{p0}$. Otherwise, for $j>0$, $E_{pj}$ and $E_{p(j-1)}$  are described by the Compton relation formula given by
\begin{equation}
	E_{pj} = \frac{E_{p(j-1)}}{1+ E_{p(j-1)}\left(1-\cos\theta_{pj}\right)}\quad \left(1<j\le \eta\right),
 \label{eqn:A2}
\end{equation}
where $0\le\theta_{pj}\le\pi$. 

\begin{table*}
    \caption{\label{tab:2} Definitions of the matrix elements $t^{(pj)}_{mn}$, where $1\le j\le\eta$ and $\eta$ is the total number of Compton scattering events and $p=1,2$ denote photon 1 or 2. Unit momentum vectors $\vb*{\hat{k}}_{p(j-1)}$ and $\vb*{\hat{k}}_{pj}$ define the direction of incident and scatter of $\gamma_{p0}$ in its interaction with the $j$-th electron (refer to Fig.\ref{fig:1}). The unit vector $\vb*{\hat{S}}_{pj}=(x_{pj},y_{pj}, z_{pj})$, where $x^{2}_{pj}+y^{2}_{pj}+z^{2}_{pj} =1$, defines the spin direction of the $j$-th incident electron.}
    \fontsize{7}{7}
 	\begin{tabular}{rlrl}
  \toprule
    	\multicolumn{2}{c}{\textbf{(a)} Spin independent terms}	& \multicolumn{2}{c}{\textbf{(b)} Spin dependent terms}\\
\midrule
     $t^{(pj)}_{11}=$ 	&	$1+\cos^{2}\theta_{pj}+\left(E_{p(j-1)}-E_{pj}\right)\left(1-\cos\theta_{pj}\right)$ 	&$t^{(pj)}_{14}=$ &$-\left(1-\cos\theta_{pj}\right)\left(E_{p(j-1)}\cos\theta\vb*{\hat{k}}_{p(j-1)} + 	E_{pj}\vb*{\hat{k}}_{pj}\right)\cdot\vb*{\hat{S}}_{pj}$\\ \vspace{5.5pt}
		$t^{(pj)}_{12} = t^{(pj)}_{21}=$	&	$\sin^{2}\theta_{pj}$	&	$t^{(pj)}_{24}=$	&	$E_{p(j-1)}\left(1-\cos\theta_{pj}\right)\left(\vb*           {\hat{k}}_{pj}\times \vb*{\hat{k}}_{p(j-1)}\right)\cdot\left(\vb*{\hat{k}}_{p(j-1)}\times\vb*{\hat{S}}_{pj}\right)$	\\\vspace{5.5pt}
		$t^{(pj)}_{13} = t^{(pj)}_{23} = t^{(pj)}_{31} = t^{(pj)}_{32}=$		&	$0$	&	$t^{(pj)}_{34}=$	&	$E_{p(j-1)}\left(1-                          \cos\theta_{pj}\right)\left(\vb*{\hat{k}}_{p(j-1)}\times\vb*{\hat{k}}_{pj}\right)\cdot\vb*{\hat{S}}_{pj}$	\\\vspace{5.5pt}
		$t^{(pj)}_{22} = 2 - t^{(pj)}_{12}=$	&	$2-\sin^{2}\theta_{pj}$	&	$t^{(pj)}_{41}=$	&	$-\left(1-                                            \cos\theta_{pj}\right)\left(E_{pj}\cos\theta_{pj}\vb*{\hat{k}}_{pj} + E_{p(j-1)}\vb*{\hat{k}}_{p(j-1)}\right)\cdot\vb*{\hat{S}}_{pj}$	                  \\\vspace{5.5pt}
		$t^{(pj)}_{33}=$				& 	$2\cos\theta_{pj}$		&	$t^{(pj)}_{42}=$	&	$E_{pj}\left(1-\cos\theta_{pj}\right)\left(\vb*{\hat{k}}_{p(j-1)}\times\vb*{\hat{k}}_{pj}\right)\cdot\left(\vb*{\hat{k}}_{pj}\times\vb*{\hat{S}}_{pj}\right)$	\\\vspace{5.5pt}
		$t^{(pj)}_{44}=$				&	$2\cos\theta_{pj} + \left(E_{p(j-1)}-E_{pj}\right)\left(1-\cos\theta_{pj}\right)\cos\theta_{pj}$ & 	              $t^{(pj)}_{43}=$	&	$-E_{pj}\left(1-\cos\theta_{pj}\right)\left(\vb*{\hat{k}}_{p(j-1)}\times\vb*{\hat{k}}_{pj}\right)\cdot\vb*{\hat{S}}_{pj}$	\\
  \bottomrule
	\end{tabular}

\end{table*}

Stokes parameters depend on the choice of axes, and a rotation matrix $M_{pj}$ relates the Stokes parameters in one coordinate system to another. If we consider a second coordinate system rotated about the direction of photon propagation at an angle of $2\phi_{pj}$ to the right of the original, the relation is given by~\cite{McMaster1961} such that
\begin{equation}
	M_{pj}=
	\begin{bmatrix}
		1 			& 			0 		 & 	0 					& 0 \\
		0 			&  \cos2\phi_{pj}	 & \sin2\phi_{pj} 	& 0 \\
		0 			& -\sin2\phi_{pj}	 &\cos2\phi_{pj} 		& 0 \\
		0			& 			0 		 & 		0 				& 1 \\
	\end{bmatrix},
	\label{eqn:A3}
\end{equation}
where  $-\pi\le\phi_{pj}\le\pi$. (Note that when considering unpolarized electrons, it is sufficient to use only the upper-left $3\times 3$ submatrices given in Eq.(\ref{eqn:A1}) and Eq.(\ref{eqn:A3}).)
\section{Stokes vectors for polarization filters}
\label{app:B}
Stokes parameters for polarization filters, which assess two orthogonal polarization states, are defined in this method with respect to the scattering plane containing the momentum vectors ($\vb*{\Vec{k}}_{\eta-1}, \vb*{\Vec{k}}_{\eta}$), as illustrated in Fig.\ref{fig:1}. The fundamental pairs of orthogonal filters are the linear, diagonal, and circular polarization filters.

In the case of linear polarization filters, one can analyze the state of polarization orthogonal to the scattering plane with $\ket{D_{\perp}}$ and parallel to it with $\ket{D_{\parallel}}$ where
\begin{equation}
    \ket{D_{\perp}}=
	\begin{bmatrix}
		1 \\
		1 \\
		0 \\
		0 \\
	\end{bmatrix},\quad
   \ket{D_{\parallel}}=
	\begin{bmatrix}
		1 \\
		-1 \\
		0 \\
		0 \\
	\end{bmatrix}.
\end{equation}

Another set of polarization filters includes the diagonal $\ket{D_{+45}}$ and anti-diagonal $\ket{D_{-45}}$ polarization filters, which have the explicit form  
\begin{equation}
    \ket{D_{+45}}=
	\begin{bmatrix}
		1 \\
		0 \\
		1 \\
		0 \\
	\end{bmatrix},\quad
   \ket{D_{-45}}=
	\begin{bmatrix}
		1 \\
		0 \\
		-1 \\
		0 \\
	\end{bmatrix}.
\end{equation}
The final pair of polarization filters includes the right circular $\ket{D_{r}}$ and left circular $\ket{D_{l}}$ polarization filters, given by
\begin{equation}
    \ket{D_{r}}=
	\begin{bmatrix}
		1 \\
		0 \\
		0 \\
		-1 \\
	\end{bmatrix},\quad
   \ket{D_{l}}=
	\begin{bmatrix}
		1 \\
		0 \\
		0 \\
		1 \\
	\end{bmatrix}.
\end{equation}
Using a polarization-insensitive detector implies making an average measurement of two orthogonal states and summing the results. Let $\ket{I}$ represent a polarization-insensitive detector. In the context of the fundamental pairs of orthogonal filters we have
\begin{equation}
    \begin{aligned}
         \ket{I} =&~\frac{1}{2}\ket{D_{\perp}}+\frac{1}{2}\ket{D_{\parallel}},\\
         \ket{I} =&~\frac{1}{2}\ket{D_{+45}}+\frac{1}{2}\ket{D_{-45}},\\
         \ket{I} =&~\frac{1}{2}\ket{D_{r}}+\frac{1}{2}\ket{D_{l}},\\
         \ket{I} =&~	\begin{bmatrix}
		1 \\
		0 \\
		0 \\
		0 \\
	\end{bmatrix}.
    \end{aligned}
\end{equation}
\section{Matrix element definitions for 3-Compton scattering}
\label{app:C}
The Compton transition matrix elements relevant to the 3-Compton cross section have the following functional forms:
	
For $\gamma_{10}\mapsto\gamma_{11}$ 
\begin{subequations}
    \begin{equation}
        t^{(11)}_{11} = 1 + \cos^2\theta_{11} + \left(E_{10} - E_{11}\right)\left(1 - \cos\theta_{11}\right),
    \end{equation}
    \begin{equation}
        t^{(11)}_{12} = \sin^{2}\theta_{11},
    \end{equation}
\end{subequations}
where $E_{10} =1$ and 
\begin{equation}
    E_{11} = \frac{1}{2-\cos\theta_{11}};\quad (\mbox{cf. Eq.(\ref{eqn:23})}).
\end{equation}

For $\gamma_{20}\mapsto\gamma_{21}$
\begin{subequations}
    \begin{equation}
        t^{(21)}_{11} = 1 + \cos^2\theta_{21} + \left(E_{20} - E_{21}\right)\left(1 - \cos\theta_{21}\right),
    \end{equation}
    \begin{equation}
        t^{(21)}_{12} = \sin^{2}\theta_{21},
    \end{equation}
    \begin{equation}
        t^{(21)}_{33} = 2\cos\theta_{21},
    \end{equation}
\end{subequations}
where $E_{20} =1$ and 
\begin{equation}
    E_{21} = \frac{1}{2-\cos\theta_{21}};\quad (\mbox{cf. Eq.(\ref{eqn:27a})}).
\end{equation}

For $\gamma_{21}\mapsto\gamma_{22}$
\begin{subequations}
    \begin{equation}
        t^{(22)}_{11} = 1 + \cos^2\theta_{22} + \left(E_{21} - E_{22}\right)\left(1 - \cos\theta_{22}\right),
    \end{equation}
    \begin{equation}
        t^{(22)}_{12} = \sin^{2}\theta_{22},
    \end{equation}
\end{subequations}
where 
\begin{equation}
    E_{22} = \frac{E_{21}}{1+ E_{21}\left(1-\cos\theta_{22}\right)};\quad (\mbox{cf. Eq.(\ref{eqn:27b})}).
\end{equation}

\bibliographystyle{unsrt}  

\providecommand{\noopsort}[1]{}\providecommand{\singleletter}[1]{#1}%

\end{document}